\documentclass[useAMS,usenatbib,usegraphicx]{mn2e}
\usepackage{amssymb}

\title[The HST colours of pop III galaxies]{The HST colours of high-redshift population III galaxies with strong Ly$\alpha$ emission}
\author[Zackrisson et al.]{Erik Zackrisson$^{1}$\thanks{E-mail: ez@astro.su.se}, Akio K. Inoue$^{2}$, Claes-Erik Rydberg$^{1}$ and Florent Duval$^{1}$\\ 
$^{1}$Department of Astronomy, Stockholm University, Oscar Klein Center, AlbaNova, Stockholm SE-106 91, Sweden\\
$^{2}$College of General Education, Osaka Sangyo University, 3-1-1, Nakagaito, Daito, Osaka 574-8530, Japan}

\begin{document}

\date{Accepted ... Received ...; in original form ...}

\pagerange{\pageref{firstpage}--\pageref{lastpage}} \pubyear{2010}

\maketitle

\label{firstpage}

\begin{abstract}
Population III galaxies, made partly or exclusively of metal-free stars, are predicted to exist at high redshifts and may produce very strong Ly$\alpha$ emission. A substantial fraction of these Ly$\alpha$ photons are likely absorbed in the intergalactic medium at $z>6$, but recent simulations suggest that significant Ly$\alpha$ emission may be detectable up to $z\approx 8.5$, i.e. well into the reionization epoch. Here, we argue that high-redshift population III galaxies with strong Ly$\alpha$ emission can be identified in Hubble Space Telescope imaging data because of their unusual colours. We quantify this effect in some of the filters used in $Y$-band dropout searches for galaxies at $z\approx 8$ and find that population III galaxies with high Ly$\alpha$ fluxes may exhibit much bluer $J-H$ colours at $z\approx 8$--10 than any normal type of galaxy at these redshifts. This colour signature can arise even if pop III stars account for as little as $\sim 10^{-3}$--$10^{-2}$ of the stellar mass in these galaxies. Some of the anomalously blue objects reported in current $Y$-band dropout samples do in fact meet the colour criteria for Ly$\alpha$-emitting population III galaxies. 
\end{abstract}


\section{Introduction}
\label{intro}
The first population III (hereafter pop III) stars are expected to form in isolation or in small numbers within $\sim$10$^5$--$10^6 M_\odot$ dark matter halos at redshifts $z\approx 10$--60 \citep[e.g.][]{Trenti & Stiavelli}, but such stars are likely too faint to be detectable even with the James Webb Space Telescope\footnote{http://www.jwst.nasa.gov/} \citep[e.g.][]{Greif et al.}. Pop III stars may, however, also continue to form in the more massive halos hosting some of the first galaxies at $z\leq 15$ \citep[e.g.][]{Scannapieco et al.,Schneider et al.,Tornatore et al.,Johnson et al. a,Johnson et al. b,Stiavelli & Trenti,Johnson}, and this could allow their integrated light to be detected, possibly even with existing telescopes. Such pop III galaxies\footnote{here defined as galaxies which, at least over certain wavelength intervals, are dominated by the light from pop III stars} may display a number of spectral signatures that set them apart from more mundane objects \citep*[e.g.][]{Tumlinson & Shull,Tumlinson et al.,Schaerer a,Schaerer b,Inoue a,Raiter et al.,Inoue b,Zackrisson et al.}. In particular, the high effective temperatures ($\sim 10^5$K) of pop III stars imply very high fluxes of hydrogen-ionizing (Lyman continuum) photons, many of which may be converted into Ly$\alpha$ photons at a rest wavelength of 1216 \AA{} after recombination in the surrounding gas. Under idealized conditions, this implies that pop III galaxies should display very high rest frame Ly$\alpha$ equivalent widths \citep[up to $\mathrm{EW}(\mathrm{Ly}\alpha) \approx 600-4000$ \AA; e.g.][]{Raiter et al.}.

The strength of the Ly$\alpha$ line can admittedly be reduced by a number of mechanisms. Feedback from pop III stars may cause Lyman continuum photons to escape directly into the intergalactic medium (IGM), although the importance of this effect depends both on the pop III stellar initial mass function (IMF) and the star formation efficiency \citep{Johnson et al. b}. Since Ly$\alpha$ is a resonant line, Ly$\alpha$ photons can also scatter repeatedly within the interstellar medium and may eventually be destroyed by dust. Current observations suggest that the fraction of Ly$\alpha$ photons that survive increases with redshift, growing from $\lesssim 0.01$ for metal-rich galaxies in the local Universe to $\approx 0.3$ for galaxies at $z\approx 6$ \citep{Hayes et al.} -- possibly due to a lower dust content in high-redshift galaxies. Since pop III galaxies are expected to have very little dust\footnote{During early stages of evolution, pure pop III galaxies are expected to have {\it no} dust, but Ly$\alpha$ photons originating from hybrid galaxies consisting of both pop III and pop II/I stars may well be subject to extinction effects}, the fraction of Ly$\alpha$ photons that escape from such galaxies may be even higher. For galaxies in the reionization epoch ($z\gtrsim 6$), the Ly$\alpha$ flux may be further reduced by absorption in the neutral IGM. However, the claimed discovery of a Ly$\alpha$ emitter at $z\approx 8.6$ \citep{Lehnert et al.} suggests that Ly$\alpha$ detections are possible well into the reionization epoch, and radiative transfer simulations predict that outflows and patchy reionization may allow a significant fraction ($\approx 0.1$--0.5) of the Ly$\alpha$ photons to evade absorption up to this redshift \citep*{Dijkstra et al.}. Other teams have also reported a few detections of objects with high $\mathrm{EW}(\mathrm{Ly}\alpha)$ at $z>6$ \citep[e.g.][]{Stark et al.,Kashikawa et al.}. Hence, pop III galaxies may plausibly display strong Ly$\alpha$ emission even at $z>6$. This would have a pronounced effect on broadband fluxes in filters transmitting the Ly$\alpha$ line, with curious colours as a result \citep[e.g.][]{Pello & Schaerer, Richard et al.}. Here, we quantify this effect in some of the {\it Hubble Space Telescope} ({\it HST}) filters used in the study of $z\approx 8$ galaxies.

\section{Model results}
To explore the impact of the Ly$\alpha$ line on the {\it HST} fluxes, we use the Yggdrasil spectral synthesis code \citep{Zackrisson et al.} with Starburst99 Padova-AGB stellar population spectra \citep{Leitherer et al.,Vazquez & Leitherer} for metal-enriched stars (pop I/II) and \citet{Schaerer a} and \citet{Raiter et al.} stellar population spectra for pop III stars. The nebular contribution to the overall spectral energy distribution (SED) is computed using the photoionization code Cloudy \citep{Ferland et al.}, assuming a spherical geometry for the photoionized gas. The model results are publicly available from the lead author's homepage\footnote{Yggdrasil model results available at: www.astro.su.se/$\sim$ez}. We have verified that the results are consistent with those produced by the \citet{Inoue b} model, which is based on a similar machinery but assumes the nebula to be plane-parallel. 

The strength of the Ly$\alpha$ line is here regulated through $f_\mathrm{Ly\alpha}$, the fraction of Ly$\alpha$ photons that evade both extinction by dust within the galaxy and absorption in the neutral intergalactic medium (IGM)\footnote{Our $f_\mathrm{Ly\alpha}$ should not be confused with the usual definition of the Ly$\alpha$ escape fraction, as $f_\mathrm{Ly\alpha}$ is the product of Ly$\alpha$ escape fraction and the IGM transmission}. When computing the colours of pop III galaxies, we limit the discussion to $f_\mathrm{Ly\alpha}$ values in the range 0--0.5, since current simulations suggest that $f_\mathrm{Ly\alpha}>0.5$ must be exceedingly rare for galaxies in the reionization epoch \citep[e.g.][]{Dijkstra et al.}. The $f_\mathrm{Ly\alpha}=0$ case corresponds to the situation where all Ly$\alpha$ photons are absorbed and there is no Ly$\alpha$ contribution to the observed colours. Since we are here focusing on galaxies in the reionization epoch, the flux at wavelengths shortward of Ly$\alpha$ is always assumed to be zero due to IGM absorption.

Concerning the possibility of Lyman continuum leakage directly into the IGM \citep[e.g.][]{Johnson et al. b}, we will for simplicity consider only the limiting cases: either complete leakage of ionizing photons  ($f_\mathrm{esc,LyC}=1$, implying a purely stellar SED) or no leakage whatsoever ($f_\mathrm{esc,LyC}=0$, which maximizes the nebular contribution to the SED). 
\begin{figure}
\includegraphics[width=84mm]{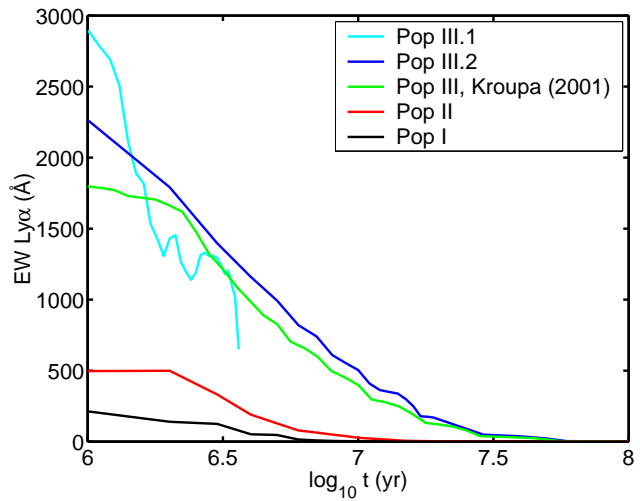}
\caption{The temporal evolution of the rest frame Ly$\alpha$ equivalent widths for instantaneous-burst populations with various metallicities and IMFs (see main text for details). To allow a comparison to previous studies, complete transmission of Ly$\alpha$ photons is assumed (i.e. $f_\mathrm{Ly\alpha}=1$). The abrupt end of the cyan line after $t\approx 3$ Myr signals the death of the least massive stars (50 $M_\odot$) in the pop III.1 scenario. During the first few Myr, the pop III models are able to reach $\mathrm{EW}(\mathrm{Ly}\alpha)\approx 1000$--3000 \AA, whereas pop I and II models are confined to $\mathrm{EW}(\mathrm{Ly\alpha})\leq 500$ \AA.} 
\label{fig1}
\end{figure}
\subsection{Ly$\alpha$ equivalent widths}
In Fig.~\ref{fig1}, we plot the predicted evolution of the intrinsic Ly$\alpha$ rest-frame equivalent width $\mathrm{EW}(\mathrm{Ly}\alpha)$ for various metallicities and IMFs, under the assumption of an instantaneous burst of star formation and $f_\mathrm{Ly\alpha}=1$.  As in \citet{Zackrisson et al.}, the pop I ($Z=0.020$) and pop II ($Z=0.0004$) models assume a \citet{Kroupa} IMF, representative for star formation in the local Universe. In the case of pop III, three different IMFs are considered: an extremely top-heavy IMF (pop III.1) with power-law slope $\alpha=2.35$ ($\mathrm{d}N/\mathrm{d}M\propto M^{-\alpha}$) throughout the mass range 50--500 $M_\odot$; a moderately top-heavy, log-normal IMF (pop III.2) with characteristic mass $M_\mathrm{c}=10 M_\odot$, dispersion $\sigma=1\ M_\odot$ and tails extending from 1--500 $M_\odot$; and finally the same \citet{Kroupa} IMF as adopted for pop I/II. 

At ages up to a few Myr, the $\mathrm{EW}(\mathrm{Ly}\alpha)$ of the pop III models are in the range $\approx 1000$--3000 \AA, which is sufficient to have a pronounced effect on {\it HST} broadband fluxes of pop III galaxies, even if just a minor fraction ($f_\mathrm{Ly\alpha}\gtrsim 0.1$) of the Ly$\alpha$ photons are transmitted to the observer. By contrast, the pop II model peaks at $\mathrm{EW}(\mathrm{Ly}\alpha)\approx 500$ \AA{} and the pop I model at $\approx 200$ \AA{}. These $\mathrm{EW}(\mathrm{Ly}\alpha)$ predictions are largely consistent with those presented by \citet{Raiter et al.}.
\begin{figure*}
\includegraphics[width=84mm]{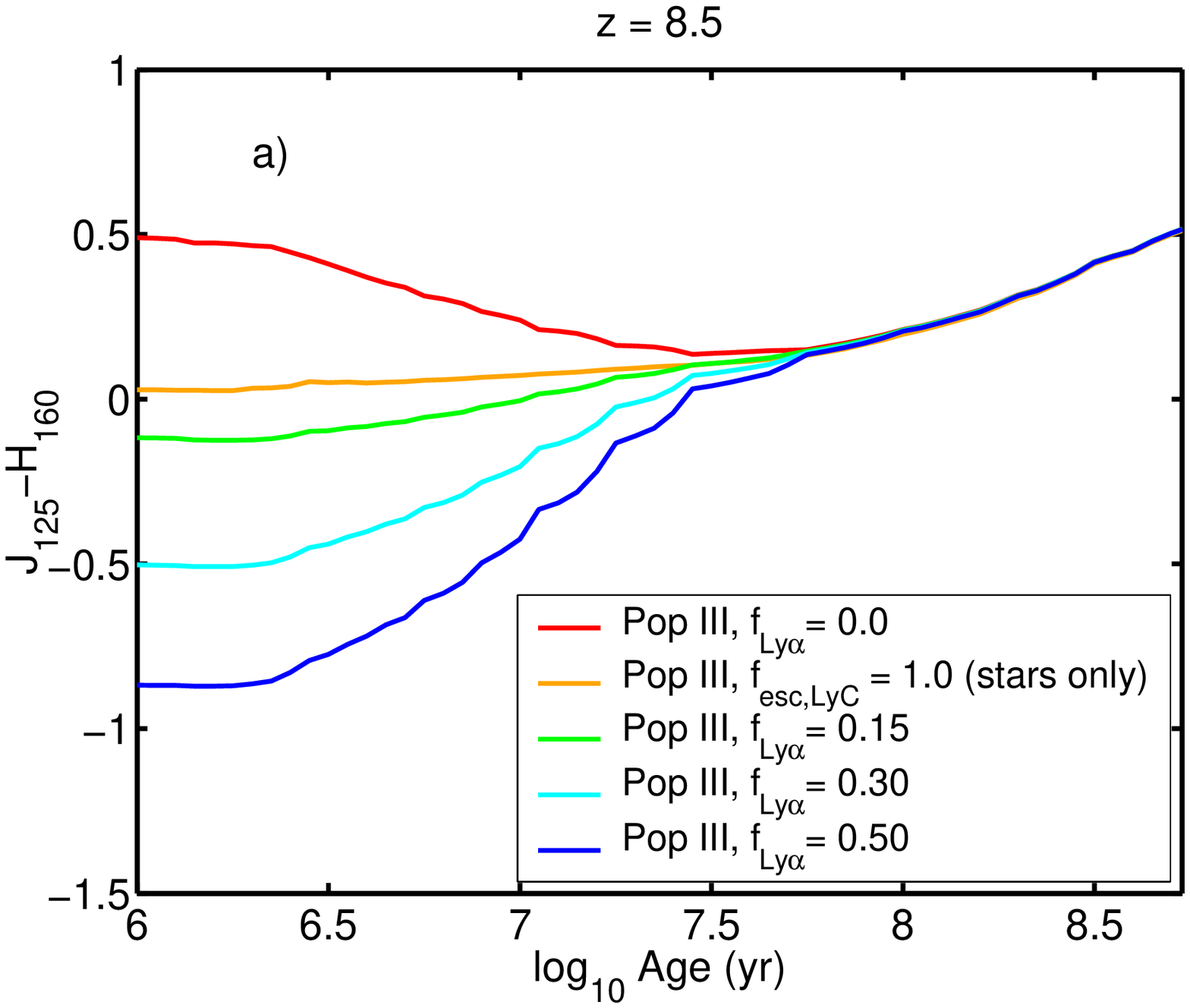}\includegraphics[width=84mm]{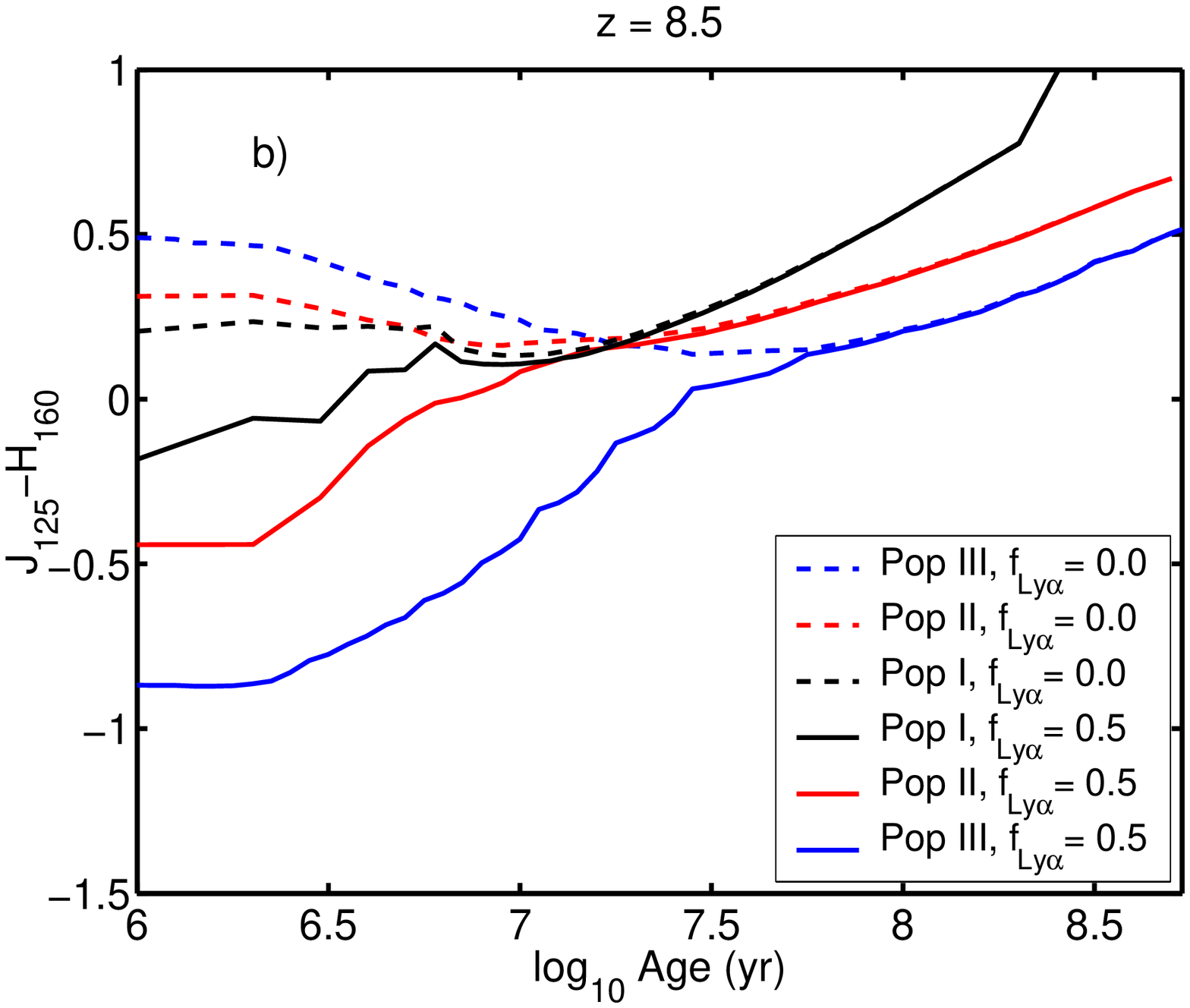}
\caption{Predicted {\it HST/WFC3} $J_{125}-H_{160}$ colours at $z=8.5$. {\bf a)} The predicted evolution for pop III galaxies with \citet{Kroupa} IMFs and different degrees of nebular and Ly$\alpha$ emission. {\bf b)} The predicted evolution for pop I, II and III galaxies with \citet{Kroupa} IMFs and nebular emission but varying of Ly$\alpha$ fractions. The bluest models at this redshift are the pop III galaxies with high Ly$\alpha$ fractions ($f_\mathrm{Ly\alpha}=0.5$; blue solid line in both panels).} 
\label{fig2}
\end{figure*}

\subsection{The colour signatures of pop III galaxies with strong Ly$\alpha$ emission}
\label{signatures}
In Fig.~\ref{fig2}, we plot the predicted evolution of the {\it HST/WFC3} $J_{125}-H_{160}$ colour for pop I, II and III galaxies at $z=8.5$ with and without residual Ly$\alpha$ emission. In the absence of any significant Ly$\alpha$ contribution in the $J_{125}$ filter, nebular emission (red line in Fig.~\ref{fig2}a) makes the $J_{125}-H_{160}$ colour redder compared to a purely stellar SED (orange line in Fig.~\ref{fig2}a). However, once a non-zero $f_\mathrm{Ly\alpha}$ is assumed (green, cyan and blue lines for $f_\mathrm{Ly\alpha}=0.1$, 0.3 and 0.5), $J_{125}-H_{160}$ become progressively bluer. A similar result was also presented by \citet{Pello & Schaerer}.

As shown in Fig.~\ref{fig2}b, young pop III galaxies with strong Ly$\alpha$ emission are moreover predicted to be bluer than both pop I and II galaxies with similar $f_\mathrm{Ly\alpha}$ at this redshift. As dust extinction likely renders $f_\mathrm{Ly\alpha}$ lower in pop I/II galaxies and further reddens the colours, it therefore seems reasonable to search for population III galaxies among the objects with the very bluest $J_{125}-H_{160}$ colours. 

The pop III models in Fig.~\ref{fig2} are all based on a \citet{Kroupa} IMF. Since more top-heavy IMFs can produce higher $\mathrm{EW}(\mathrm{Ly}\alpha)$ (see Fig.~\ref{fig1}), one would naively expect that such pop III models should give rise to even more extreme colours. However, we find this not to be the case. While the pop III IMF is important for determining the duration of the luminous phase of a pop III galaxy, the colours of pop III galaxies with \citet{Kroupa}, pop III.2 and pop III.1 IMFs are nearly identical at young ages, despite different $\mathrm{EW}(\mathrm{Ly}\alpha)$. This is because the higher Lyman continuum flux associated with a more top-heavy IMF boosts the strength of the Ly$\alpha$ emission line {\it and} the relative strength of nebular continuum directly longward of Ly$\alpha$. In these filters, the two effects balance each other to make the overall colours fairly insensitive to the overall IMF.

\begin{figure}
\includegraphics[width=84mm]{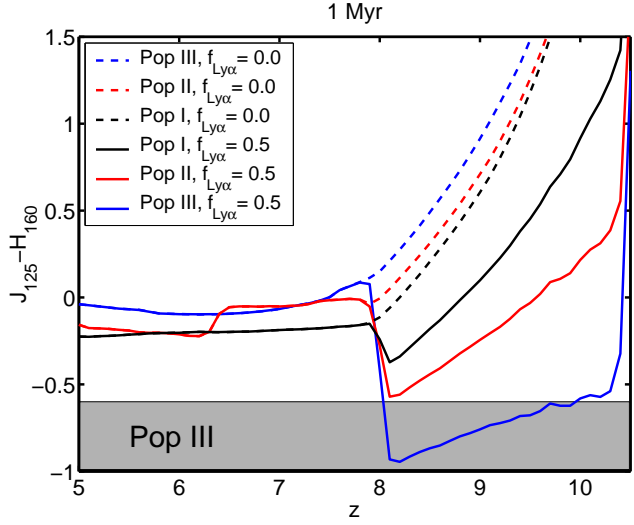}
\caption{The redshift evolution of the {\it HST/WFC3} $J_{125}-H_{160}$ colour for 1 Myr old galaxies with various degrees of nebular and Ly$\alpha$ emission. The models used are the same as in Fig.~\ref{fig2}b. Among reionization-epoch galaxies, only pop III galaxies with strong Ly$\alpha$ emission (blue solid line) display $J_{125}-H_{160}\leq -0.6$, which makes objects in this colour range (gray region) excellent pop III candidates throughout the redshift interval $z\approx 8$--10.} 
\label{fig3}
\end{figure}

In Fig.~\ref{fig3}, we show the predicted redshift evolution of the $J_{125}-H_{160}$ colour for newborn (1 Myr old) galaxies of various metallicities and $f_\mathrm{Ly\alpha}$ (the same models as in Fig.~\ref{fig2}b). Throughout the redshift interval $z\approx 8$--10, pop III galaxies with significant Ly$\alpha$ emission ($f_\mathrm{Ly\alpha}=0.5$) display $J_{125}-H_{160}\leq -0.6$. Under the assumption that no other galaxies display $f_\mathrm{Ly\alpha}>0.5$, pop III objects are the only normal reionization-epoch ($z>6$) galaxies that are expected to display such colours. Objects with $J_{125}-H_{160}\leq -0.6$ (gray region in Fig.~\ref{fig3}) therefore represent excellent population III galaxy candidates. An age of 1 Myr is adopted in this plot since this produces the  bluest colours possible for pop III, II and I galaxies (see Fig.~\ref{fig2}b). Higher ages would imply redder colours, as demonstrated in Fig.~\ref{fig2} for an instantaneous burst, but pop III galaxies with constant star formation rates can in principle retain their very blue colours ($J_{125}-H_{160}\leq -0.6$) for up to $\sim 10^8$ yr. 

\begin{figure}
\includegraphics[width=84mm]{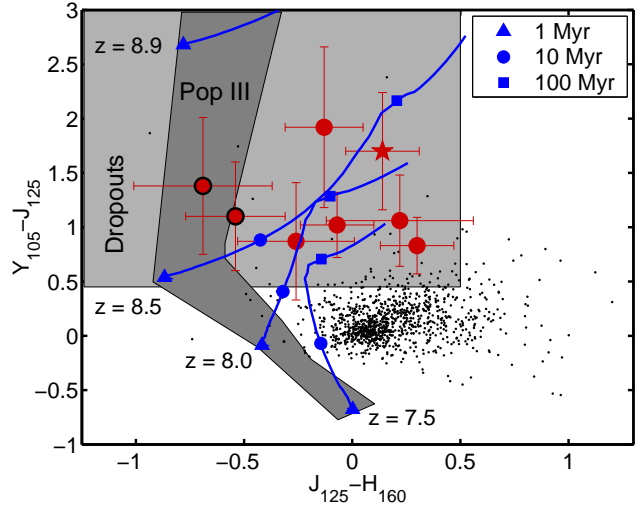}
\caption{The location of $z\approx 8$ pop III galaxies in the $J_{125}-H_{160}$ vs. $Y_{105}-J_{125}$ plane. Blue lines represent pop III, instantaneous burst models with $f_\mathrm{Ly\alpha}=0.5$ at $z=7.5$, 8.0, 8.5 and 8.9. Along these tracks, triangles, circles and squares indicate ages of 1, 10  and 100 Myr. The light gray region indicates the $Y$-band dropout criteria used by \citet{Bouwens et al.} to select $z\approx 8$ galaxies. The dark gray region represents the region expected to be populated exclusively by pop III galaxies with strong Ly$\alpha$ emission at $z=7$--9. Red markers represent the objects in the \citet{Taniguchi et al.} compilation of $z\approx 8$ galaxies. Two objects in this sample (red symbols surrounded by black circles) are potential pop III galaxy candidates. The red star marks the position of the \citet{Lehnert et al.} spectroscopically confirmed $z\approx 8.6$ Ly$\alpha$-emitter. To indicate the typical colours of objects in deep {\it HST} fields, the objects in the \citet{Cameron et al.} HUDF09 source catalog have been included as black dots.}
\label{fig4}
\end{figure}

Intriguingly, reionization-epoch objects with colours expected for young pop III galaxies with Ly$\alpha$ emission may already have been detected. The \citet*{Taniguchi et al.} compilation of $z\approx 8$ objects contains one objects at $J_{125}-H_{160}\leq -0.6$ and one at $J_{125}-H_{160}\leq -0.5$. In Fig.~\ref{fig4}, we compare the predicted $J_{125}-H_{160}$ and $Y_{105}-J_{110}$ colours of Ly$\alpha$-emitting pop III galaxies at $z\approx 8$ to those of the \citet{Taniguchi et al.} sample and the $Y_{105}$-band dropout criteria (light gray region) typically used to select $z\approx 8$ targets from multiband {\it HST} data \citep{Bouwens et al.}. This diagram also includes a dark gray, pop III galaxy zone. The extent of this area has been derived by studying the evolution of all pop I, II, III galaxies at $z=7$--9 in our model grid and identifying the region in which only pop III galaxies with strong Ly$\alpha$ emission should appear. Indeed, the two bluest objects (red markers with black rings) in $J_{125}-H_{160}$ are broadly consistent with the colour criteria for pop III galaxies, at a likely redshift of $z\approx 8.1$--8.9.  While not included in this plot, we note that the sample of \citet{McLure et al.} includes two more objects with $J_{125}-H_{160}\lesssim -0.5$ (HUDF-2641 and ERS-9041, in their notation).

\citet{Dunlop et al.} and \citet{Wilkins et al.} caution that spurious detections of anomalously blue objects such as these are expected close to the detection thresholds. Follow-up spectroscopy is therefore required to unravel the exact nature of these pop III candidates. If their colours are approximately correct and indeed due to Ly$\alpha$, our models predict these objects to display rest-frame $\mathrm{EW}(\mathrm{Ly}\alpha)\gtrsim 200$ \AA.

Fig.~\ref{fig4} also reveals that pop III galaxies with strong Ly$\alpha$ emission require higher redshifts than normal galaxies to enter a given dropout-selected sample. \citet{Bouwens et al.} adopt a  $Y_{105}$-band dropout criteria of $Y_{105}-J_{125}>0.45$ and $J_{105}-H_{160}<0.5$ for their $z\sim 8$ sample (light gray region), and estimate that this will select objects in the $z\approx 7$--9.2 interval. However, since the $Y_{105}$ and $J_{125}$ filters partly overlap, a strong Ly$\alpha$ line at $z\approx 8$ can affect both fluxes and prevent the $Y_{105}-J_{125}>0.45$ condition from applying. Our models for young pop III, $f_\mathrm{Ly\alpha}=0.5$ galaxies suggest that such objects can escape the dropout criterion until $z\geq 8.1$ and then remain a $Y_{105}$-band dropout (without becoming a $J_{125}$ band dropout) as long as $z\leq 10.4$. At $7.0<z<8.1$, these objects would instead enter the \citet{Bouwens et al.} $z_{850}$-band dropout sample (assumed to contain objects in the range $z\approx 6$--7.5). Hence, dropout selection schemes of this type may lead the redshifts of strong Ly$\alpha$-emitters to be underestimated by up to $\Delta(z)\approx 0.5$--1. 

\section{Discussion}
\label{discussion}
The results in Sect.~\ref{signatures} are valid for pure pop III galaxies, i.e. objects consisting of pop III stars only. However, the detection of such galaxies at $z\approx 8.5$ in the deepest {\it HST} fields ($J_{125}\leq 29.9$) would require stellar population masses of $\gtrsim 10^5 M_\odot$ for a pop III.1 IMF and $\gtrsim 10^6 M_\odot$ for a \citet{Kroupa} IMF, assuming $f_{Ly\alpha}\leq 0.5$. In the models of \citet{Trenti et al.} and \citet{Stiavelli & Trenti}, the most massive pure pop III galaxies at $z>7$ have halo masses of $M\sim 10^8\ M_\odot$ and baryonic masses of $M\sim 10^7\ M_\odot$. Hence, the star formation efficiency (the fraction of the total halo gas mass converted into stars) would need to be very high ($\epsilon\gtrsim 0.01$--0.1) to make such objects detectable in current data.  

Galaxies in which pop III and pop I/II stars form in parallel can attain much higher total masses, but any pop III spectral signature may at the same time be diluted beyond recognition. While \citet{Jimenez & Haiman} interpret the Ly$\alpha$ properties of galaxies at $z\approx 3$--4 as due to a very large mass fraction ($\gtrsim 0.1$) in pop III stars, simulations tend to favour a much smaller contribution \citep[e.g.][]{Salvaterra et al.}. To explore the requirements for a hybrid galaxy to display the pop III spectral signatures discussed in this paper, we define a mass ratio $f_\mathrm{pop\ III}$ of pop III stars within a galaxy: $f_\mathrm{pop\ III}=M_\mathrm{pop\ III}/(M_\mathrm{pop\ III}+M_\mathrm{pop\ I/II})$, where the different masses represent the gas masses converted into stars prior to the age at which the galaxy is observed. Since a young burst of pop III stars has a rest-frame UV stellar $M/L$ ratio that is much smaller than that of its pop I/II host galaxy, the pop III component can dominate the UV flux even if its mass is significantly lower. 

Obviously, the $f_\mathrm{pop\ III}$ ratio required to produce extremely blue colours due to Ly$\alpha$ domination depends on many factors, including the pop III IMF and the star formation history of the pop I/II component. However, assuming a 1 Myr old pop III burst located in a 300 Myr old pop I galaxy with a constant star formation rate (both having $f_\mathrm{Ly\alpha}=0.5$) at $z=8.5$, we find that extremely blue colours ($J_{105}-H_{160}\leq-0.6$) can be produced for $f_\mathrm{pop\ III}\geq 7\times10^{-2}$ if the pop III IMF is similar to that of \citet{Kroupa}, for $f_\mathrm{pop\ III}\geq 1\times10^{-2}$ in the case of a pop III.2 IMF and for $f_\mathrm{pop\ III}\geq 4\times 10^{-3}$ in the case of a pop III.1 IMF. If the pop III burst takes place in a more passively evolving pop I/II galaxy (modelled as in instantaneous burst), the ratio can be as low as $f_\mathrm{pop\ III}\approx 5\times 10^{-3}$ even for a \citet{Kroupa} pop III IMF and $\approx 3\times 10^{-4}$ for a pop III.1 IMF.

The models presented in this paper assume the stellar IMF to be fully sampled. This assumption breaks down in systems containing small numbers of stars, with drastically different colour predictions as a result \citep[e.g.][]{Cerviño & Valls-Gabaud}. While pure pop III galaxies are expected to be low-mass systems with limited numbers of stars, IMF sampling effects are unlikely to jeoparize colour signatures dominated by strong Ly$\alpha$ fluxes. As discussed in Sect.~\ref{signatures}, our colour predictions are very similar for the three pop III IMFs considered, and this holds as long as nebular emission dominates over direct star light. To have any significant impact on the colours, IMF sampling effects would have to result in a situation where almost no pop III stars with mass $\gtrsim 10\ M_\odot$ are present, as this would imply very little ionizing radiation. This is impossible for the two top-heavy IMFs considered (pop III.1 and III.2), but may in principle occur for the \citet{Kroupa} IMF. However, this would also dramatically reduce the luminosity of such systems. We estimate that a stellar population mass of $\gtrsim 10^{7-8}\ M_\odot$ would be required to make such pop III galaxies detectable in current {\it HST} data at $z\approx 8.5$. This is orders of magnitude higher than the mass scale where IMF sampling is likely to affect photoionization calculations \citep{Villaverde et al.}. Hence, for pop III galaxies in the luminosity range relevant for current observations, IMF sampling is not likely to be an issue.

Finally, we note that while no normal pop I/II galaxies are expected to to enter the $J_{110}-H_{125}<-0.6$ pop III region at $z>5$ (see Fig.~\ref{fig3}), it is conceivable that other unusual high-redshift objects could display similar colours. Gas cooling \citep[e.g.][]{Dijkstra} or fast accretion shocks \citep{Dopita et al.} in massive halos may also produce strong Ly$\alpha$ emission with high equivalent widths, and so can accreting black holes \citep[e.g.][]{Haiman & Rees,Johnson et al. c}. Since the identification of any of these mechanisms at $z\gtrsim 8$ would be a remarkable discovery in its own right, follow-up studies of objects with pop III-like colour signatures are definitely justified.

\section{Acknowledgements}
E.Z acknowledges funding from the Swedish National Space Board and the Swedish Research Council. A.K.I. acknowledges funding from the Ministry of Education, Culture, Sports, Science, and Technology (MEXT) of Japan (KAKENHI: 23684010). C-E.R. acknowledges funding from the Swedish National Space Board and the Royal Swedish Academy of Sciences.

\end{document}